\documentstyle[12pt,epsf]{article}
\begin{document}
\def \beq{\begin{equation}}
\def \eeq{\end{equation}}
\def \g{{\rm GeV}}
\rightline{DOE/ER/40561-210-INT95-17-01}
\rightline{EFI-95-41}
\rightline{hep-ph/9507375}
\vspace{0.5in}
\centerline{\bf IMPACT OF ATOMIC PARITY VIOLATION MEASUREMENTS}
\centerline{\bf ON PRECISION ELECTROWEAK PHYSICS
\footnote{To be submitted as a Brief Report to Phys.~Rev.~D.}}
\vspace{0.5in}
\centerline{\it Jonathan L. Rosner}
\centerline{\it Institute for Nuclear Theory}
\centerline{\it University of Washington, Seattle, WA 98195}
\bigskip
\centerline{and}
\bigskip
\centerline{\it Enrico Fermi Institute and Department of Physics}
\centerline{\it University of Chicago, Chicago, IL 60637
\footnote{Permanent address.}}
\bigskip

\centerline{\bf ABSTRACT}
\medskip
\begin{quote}
The impact of atomic parity violation experiments on determination of the weak
mixing parameter $\sin^2 \theta$ and the Peskin-Takeuchi parameters $S$ and $T$
is reassessed in the light of recent electroweak measurements at LEP, SLAC, and
Fermilab.  Since the weak charge $Q_W$ provides unique information on $S$, its
determination with a factor of four better accuracy than present levels can
have a noticeable effect on global fits.  However, the measurement of $\Delta
Q_W / Q_W$ for two different isotopes provides primarily information on $\sin^2
\theta$. To specify this quantity to an accuracy of $\pm 0.0004$, comparable to
that now provided by other electroweak experiments, one would have to determine
$\Delta Q_W/Q_W$ in cesium to about 0.1\% of its value, with comparable demands
for other nuclei.  The relative merits of absolute measurements of $Q_W$ and
isotope ratios for discovering effects of new gauge bosons are noted briefly.
\end{quote}
\newpage

About five years ago it was recognized \cite{MR,PS} that the precise knowledge
of the $Z$ boson mass then becoming available would lead to a nearly unique
prediction for atomic parity-violating effects in a wide range of nuclei,
independently of standard model parameters such as the top quark mass $m_t$,
the Higgs boson mass $M_H$, or the weak mixing angle $\sin^2 \theta$.  Thus any
deviations of the weak charge $Q_W$ measured in such experiments from
theoretical expectations would have to be ascribed to physics beyond the
standard model.

A description of effects of new physics on electroweak gauge boson propagators
(the so-called ``oblique'' corrections) was introduced by Peskin and Takeuchi
\cite{PT} in terms of parameters called $S,~T$, and $U$.  The parameter $S$
describes wave-function renormalization effects, $T$ describes violations of a
``custodial SU(2)'' symmetry such as arise from the large $t-b$ mass
difference, and $U$ describes differences between wave-function
renormalizations of the $W$ and $Z$ propagators. (The only electroweak
observable sensitive to $U$ is the $W$ mass.)

In terms of these parameters, the measurement \cite{CW} of atomic parity
violation (APV) in cesium to an experimental accuracy of 2.2\% (for which the
theoretical interpretation \cite{Csth}, standing at a 1.2\% level, is most
precise) was found to constrain $S$ almost exclusively, with the $T$ dependence
nearly cancelling.  The $S$-dependence of the cesium measurement provided a
useful constraint on global fits of electroweak parameters in terms of the
Peskin-Takeuchi variables.

Since the original analysis \cite{MR}, precise electroweak data have been
obtained in many experiments at LEP \cite{Schaile}; in the measurement of the
asymmetry for polarized-electron positron annihilation at the $Z$ at SLAC
\cite{SLC}; and in the discovery of the top quark \cite{top}, the more precise
measurement of the $W$ mass \cite{Wmass}, and in the analysis of
neutral-current deep inelastic neutrino scattering \cite{CCFR} at Fermilab.
These results, when combined in a global fit, provide very strong constraints
on $\bar x$ and on the Peskin-Takeuchi parameters.  Concurrently, precise
measurements of atomic parity-violation effects have appeared in a number of
nuclei, including a 2\% measurement in bismuth \cite{Bi}, a 1\% measurement in
lead \cite{Pb}, and 1\% and 3\% measurements in thallium \cite{TlS, TlO}.  The
theoretical calculations for these effects are at levels of about 11\% for
bismuth, 8\% for lead, and 3\% for thallium.

It is the purpose of this Brief Report to indicate the precision to which APV
experiments (and the accompanying theoretical calculations) have to specify
$Q_W$ in order to have a significant impact on present global fits to
electroweak data.

We begin with a brief review of notation and formalism \cite{DPF}.  We then
specify the data germane to our fit and perform an analysis including APV data
at their present level of precision and with hypothetical errors reduced by an
appropriate factor.  We then discuss the effects of measurements of isotope
ratios, and conclude with remarks on the relative merits of absolute
measurements of $Q_W$ and isotope ratios for discovering effects of new gauge
bosons.

The low-energy limits of $W$ and $Z$ exchange are described by
\beq \label{eqn:lo}
\frac{G_F}{\sqrt{2}} = \frac{g^2}{8M_W^2} ~~~,~~~
\frac{G_F}{\sqrt{2}} \rho = \frac{g^2 +{g'}^2}{8M_Z^2} ~~~,
\eeq
where $G_F$ is the Fermi constant, $g = e/\sin \theta$ and $g' = e/\cos \theta$
are SU(2) and U(1) coupling constants, $e$ is the proton charge, and $\theta$
is the weak mixing angle. The parameter $\rho$, which receives contributions
from quark loops to $W$ and $Z$ self-energies, is dominated by the top
\cite{Tini}:
\beq \label{eqn:rho}
\rho \simeq 1 + \frac{3G_F m_t^2}{8 \pi^2 \sqrt{2}} ~~~,
\eeq
Consequently, if we define $\theta$ by means of the precise measurement
at LEP of $M_Z$,
\beq
M_Z^2 = \frac{\pi \alpha}{\sqrt{2} G_F \rho \sin^2 \theta \cos^2 \theta}
{}~~~,
\eeq
then $\theta$ will depend on $m_t$, and so will
\beq
M_W^2 = \frac{\pi \alpha}{\sqrt{2} G_F \sin^2 \theta}~~~.
\eeq

Here one must use the value of $\alpha$ appropriate to the electroweak scale
\cite{alpha}; we take $\alpha^{-1}(M_Z) = 128.9 \pm 0.1$.

The Higgs boson also affects the parameter $\rho$ through loop diagrams. It is
convenient to express contributions to $\rho$ in terms of deviations of the top
quark and Higgs boson masses from nominal values.  For $m_t = 175$ GeV, $M_H =
300$ GeV, the measured value of $M_Z$ leads to a nominal expected value of
$\sin^2 \theta_{\rm eff} = 0.2315.$  In what follows we shall interpret the
effective value of $\sin^2 \theta$ as that measured via leptonic vector and
axial-vector couplings: $\sin^2 \theta_{\rm eff} \equiv (1/4)(1 -
[g_V^{\ell}/g_A^{\ell}])$.  We have corrected the nominal value of $\sin^2
\theta_{\rm \overline{MS}} \equiv \hat s^2$ as quoted by DeGrassi, Kniehl, and
Sirlin \cite{DKS} for the difference \cite{GS} $\sin^2 \theta_{\rm eff} - \hat
s^2 = 0.0003$ and for the recent change in the evaluation of $\alpha(M_Z)$
\cite{alpha}.

Defining the parameter $T$ by $\Delta \rho \equiv \alpha T$, we find
\begin{equation}
T \simeq \frac{3}{16 \pi \sin^2 \theta} \left[ \frac{m_t^2 - (175
{}~{\rm GeV})^2}{M_W^2} \right] - \frac{3}{8 \pi \cos^2 \theta}
\ln \frac{M_H}{300~{\rm GeV}} ~~~.
\end{equation}
The weak mixing angle $\theta$, the $W$ mass, and other electroweak observables
depend on $m_t$ and $M_H$.

The weak charge-changing and neutral-current interactions are probed under a
number of different conditions, corresponding to different values of momentum
transfer.  For example, muon decay occurs at momentum transfers small with
respect to $M_W$, while the decay of a $Z$ into fermion-antifermion pairs
imparts a momentum of nearly $M_Z/2$ to each member of the pair. Small
``oblique'' corrections \cite{PT}, logarithmic in $m_t$ and $M_H$, arise from
contributions of new particles to the photon, $W$, and $Z$ propagators. Other
(smaller) ``direct'' radiative corrections are important in calcuating actual
values of observables.

We may then replace (\ref{eqn:lo}) by
\beq
\frac{G_F}{\sqrt{2}} = \frac{g^2}{8 M_W^2} \left( 1 + \frac{\alpha S_W}{4
\sin^2 \theta} \right)~~~,~~~
\frac{G_F \rho}{\sqrt{2}} = \frac{g^2 + {g'}^2}{8M_Z^2} \left( 1 + \frac{\alpha
S_Z}{4 \sin^2 \theta \cos^2 \theta} \right)~~~,
\eeq
where $S_W$ and $S_Z$ are coefficients representing variation with momentum
transfer. Together with $T$, they express a wide variety of electroweak
observables in terms of quantities sensitive to new physics.  The
Peskin-Takeuchi variable $U$ is equal to $S_W - S_Z$, while $S \equiv S_Z$.

Expressing the ``new physics'' effects in terms of deviations from nominal
values of top quark and Higgs boson masses, we have the expression for $T$
written above, while contributions of Higgs bosons and of possible new fermions
$U$ and $D$ with electromagnetic charges $Q_U$ and $Q_D$ to $S_W$ and $S_Z$
are \cite{KL}
\beq \label{eqn:sz}
S_Z = \frac{1}{6 \pi} \left [
\ln \frac{M_H}{300~\g} + \sum N_C \left ( 1 - 4 \overline Q \ln
\frac{m_U}{m_D} \right ) \right ] ~~~,
\eeq
\beq \label{eqn:sw}
S_W = \frac{1}{6 \pi} \left [
\ln \frac{M_H}{300 ~\g} + \sum N_C \left ( 1 - 4 Q_D \ln \frac{m_U}{m_D}
\right ) \right ]~~.
\eeq
The expressions for $S_W$ and $S_Z$ are written for doublets of fermions with
$N_C$ colors and $m_U \geq m_D \gg m_Z$, while $\overline Q \equiv (Q_U + Q_D )
/2$. The sums are taken over all doublets of new fermions. In the limit $m_U =
m_D$, one has equal contributions to $S_W$ and $S_Z$. For a single Higgs boson
and a single heavy top quark, Eqs.~(\ref{eqn:sz}) and (\ref{eqn:sw}) become
\beq
S_Z = \frac{1}{6 \pi} \left [ \ln \frac{M_H}{300~\g} - 2 \ln
\frac{m_t}{175~\g} \right ] ~~;~~~
S_W = \frac{1}{6 \pi} \left [ \ln \frac{M_H}{300~\g} + 4 \ln
\frac{m_t}{175~\g} \right ] ~~.
\eeq

We now list the electroweak observables used in our fit.

Recent direct $W$ mass measurements, in GeV, include
$79.92 \pm 0.39$ \cite{OldCDFW},
$80.35 \pm 0.37$ \cite{UA2W}, and
$80.41 \pm 0.18$ \cite{Wmass},
with average $80.33 \pm 0.15$.  Data \cite{CCFR,CDHS,CHARM} on the ratio $R_\nu
\equiv \sigma(\nu N \to \nu + \ldots)/\sigma(\nu N \to \mu^- + \ldots)$ lead to
information on $\rho^2$ times a function of $\sin^2 \theta$ roughly equivalent
to the constraint $M_W = 80.27 \pm 0.26$ GeV.

Measured $Z$ parameters \cite{Schaile} include $M_Z = 91.1887 \pm 0.0022~{\rm
GeV}$, $\Gamma_Z = 2.4971 \pm 0.0033~{\rm GeV}$, $\sigma_h^0 = 41.492 \pm
0.081$ nb (the hadron production cross section), and $R_\ell \equiv \Gamma_{\rm
hadrons}/\Gamma_{\rm leptons} = 20.800 \pm 0.035$, which may be combined to
obtain the $Z$ leptonic width $\Gamma_{\ell\ell}(Z) = 83.94 \pm 0.13$ MeV.
Leptonic asymmetries include the forward-backward asymmetry parameter
$A_{FB}^{\ell}$ leading to a value $\sin^2 \theta_{\rm eff}  = 0.23096 \pm
0.00073$, and independent determinations from the parameters $A_\tau \to \sin^2
\theta_{\rm eff} = 0.2324 \pm 0.0010$ and $A_e \to \sin^2 \theta_{\rm eff}
= 0.2328 \pm 0.0011$. The last three values may be combined to yield $\sin^2
\theta_{\rm eff} = 0.23176 \pm 0.00052$. [We do not use asymmetries as
measured in decays of $Z$ to $b \bar b$ (which may reflect additional
new-physics effects \cite{Zbb}), to $c \bar c$ (which are of limited weight
because of large errors), or to light quarks (for which interpretations are
more model-dependent).]  This last result is to be compared with that based on
the left-right asymmetry parameter $A_{LR}$ measured with polarized electrons
at SLC \cite{SLC}: $\sin^2 \theta_{\rm eff} = 0.2305 \pm 0.0005$.

Parity violation in atoms, stemming from the interference of $Z$ and photon
exchanges between the electrons and the nucleus, provides further information
on electroweak couplings.  The most precise constraint at present arises from
the measurement of the {\it weak charge} (the coherent vector coupling of the
$Z$ to the nucleus), $Q_W = \rho(Z - N - 4 Z \sin^2 \theta)$, in atomic cesium
\cite{CW}, with the result $Q_W({\rm Cs}) = -71.04 \pm 1.58 \pm 0.88$.  The
first error is experimental, while the second is theoretical \cite{Csth}.  The
prediction \cite{MR} $Q_W({\rm Cs}) = -73.20 \pm 0.13$ is insensitive to
standard-model parameters \cite{MR,PS}; discrepancies are good indications of
new physics (such as exchange of an extra $Z$ boson).  Recently the weak charge
has also been measured in atomic thallium.  The Seattle group \cite{TlS}
obtains $Q_W({\rm Tl}) = -114.2 \pm 3.8$, to be compared with the theoretical
estimate \cite{Tlth,PSBL} $Q_W = -116.8$.  From information presented by the
Oxford group \cite{TlO} we deduce their value of $Q_W({\rm Tl})$ to be $-120.5
\pm 5.3$.

We have performed a fit to the electroweak observables listed in Table 1.  The
``nominal'' values (including \cite{DKS} $\sin^2 \theta_{\rm eff} = 0.2315$)
are calculated for $m_t = 175$ GeV and $M_H = 300$ GeV.  We use $\Gamma_{\ell
\ell}(Z)$, even though it is a derived quantity, because it has little
correlation with other variables in our fit.  It is mainly sensitive to the
axial-vector coupling $g_A^\ell$, while asymmetries are mainly sensitive to
$g_V^\ell$.  We also omit the total width $\Gamma_{\rm tot}(Z)$ from the fit,
since it is highly correlated with  $\Gamma_{\ell \ell}(Z)$ and mainly provides
information on the value of the strong fine-structure constant $\alpha_s$.
With $\alpha_s = 0.12 \pm 0.01$, the observed total $Z$ width is consistent
with predictions.  The partial width $\Gamma(Z \to b \bar b)$ is the subject of
several discussions of new physics \cite{Zbb} which we do not address here.

Each observable in Table 1 specifies a band in the $S - T$ plane with different
slope, as seen from the ratios of coefficients of $S$ and $T$.  Parity
violation in atomic cesium and thallium is sensitive almost entirely to
$S$ \cite{MR,PS}. The impact of $\sin^2 \theta_{\rm eff}$ determinations on $S$
is considerable. The leptonic width of the $Z$ is sensitive primarily to $T$.
The $W$ mass specifies a band of intermediate slope in the $S-T$ plane; here we
assume $S_W = S_Z$.  Strictly speaking, the ratio $R_\nu$ specifies a band
with slightly more $T$ and less $S$ dependence than $M_W$ \cite{MR,PT};
we have ignored this difference here.

The resulting constraints on $S$ and $T$ are shown in Fig.~1(a).  A top quark
mass of $180 \pm 12$ $\g$ (the CDF and D0 average) is compatible with all Higgs
boson masses between 100 and 1000 $\g$, as seen by the curved lines
intersecting the error ellipses.  Independently of the standard model
predictions, values of $S$ between $-0.5$ and $0.3$ are permitted at the 90\%
confidence level.  This is to be compared with the determinations $S = -2.7 \pm
2.3$ \cite{CW}, $S = -2.2 \pm 3.2$ \cite{TlS}, and $S = 3.2 \pm 4.5$ \cite{TlO}
based on cesium and the two recent thallium experiments. Averaging, we find $S
= -1.7 \pm 1.7$. It is clear that the value of $S$ is now known much more
precisely than specified by the APV experiments.  Omission of the APV data
(the first three lines) in Table 1 in the fit shifts the ellipses by
$\Delta S = 0.017,~\Delta T = 0.016$ without affecting their sizes noticeably.

What improvement in accuracy of the APV experiments would begin to have an
impact on the fits?  Since the 90\% confidence level limits on $S$ are of
order $\pm 0.4$, one should ask for a factor of about 4 improvement in the
combined error on $Q_W$ from cesium and thallium. The effect of reducing the
total errors in each experiment by a factor of 4 while keeping the same central
values is shown in Fig.~1(b).  The standard model predictions now graze the
edge of the 90\% c.l. ellipse.

\begin{table}
\begin{center}
\caption{Electroweak observables described in fit}
\medskip
\begin{tabular}{c c c} \hline
Quantity        &   Experimental   &   Theoretical \\
                &      value       &    value      \\ \hline
$Q_W$ (Cs)      & $-71.0 \pm 1.8^{~a)} $  &  $ -73.2^{~b)} - 0.80S - 0.005T$\\
$Q_W$ (Tl)      & $-114.2 \pm 3.8^{~c)} $ &  $ -116.8^{~d)} -1.17S - 0.06T$ \\
$Q_W$ (Tl)      & $-120.5 \pm 5.3^{~e)} $ &  $ -116.8^{~d)} -1.17S - 0.06T$ \\
$M_W$ (GeV)     & $80.31 \pm 0.14^{~f)}$  & $80.35^{~g)} -0.29S + 0.45T$ \\
$\Gamma_{\ell\ell}(Z)$ (MeV) & $83.94 \pm 0.13^{~h)}$ & $83.90 -0.18S
+ 0.78T$ \\
$\sin^2 \theta_{\rm eff}$ & $0.23176 \pm 0.00052^{~i)}$ & $0.2315^{~j)}
 + 0.0036S - 0.0026T$ \\
$\sin^2 \theta_{\rm eff}$ & $0.2305 \pm 0.0005^{~k)}$ & $0.2315^{~j)} + 0.0036S
- 0.0026T$ \\
\end{tabular}
\end{center}
\leftline{$^{a)}$ {\small Weak charge in cesium \cite{CW}}}
\leftline{$^{b)}$ {\small Calculation \cite{MR} incorporating
atomic physics corrections \cite{Csth}}}
\leftline{$^{c)}$ {\small Weak charge in thallium \cite{TlS}}}
\leftline{$^{d)}$ {\small Calculation \cite{PSBL} incorporating
atomic physics corrections \cite{Tlth}}}
\leftline{$^{e)}$ {\small Weak charge in thallium \cite{TlO}}}
\leftline{$^{f)}$ {\small Average of direct measurements
and indirect information}}
\leftline{{\small \quad from neutral/charged current ratio in
deep inelastic neutrino scattering \cite{CCFR,CDHS,CHARM}}}
\leftline{$^{g)}$ {\small Including perturbative QCD corrections \cite{DKS}}}
\leftline{$^{h)}$ {\small LEP average as of May, 1995 \cite{Schaile}}}
\leftline{$^{i)}$ {\small From asymmetries at LEP \cite{Schaile}}}
\leftline{$^{j)}$ {\small As calculated \cite{DKS} with correction for
relation between $\sin^2 \theta_{\rm eff}$ and $\hat s^2$ \cite{GS}}}
\leftline{$^{k)}$ {\small From left-right asymmetry in annihilations at
SLC \cite{SLC}}}
\end{table}

The comparison of $Q_W$ for more than one isotope can provide electroweak
information in which atomic physics corrections play a much less significant
role \cite{ratio}.  One can measure the ratio of the difference for two
different isotopes, $\Delta Q_W \equiv Q_W(N_1) - Q_W(N_2)$, with respect to an
average value $\bar Q_W \equiv [Q_W(N_1) + Q_W(N_2)]/2$ for the two. Since $Q_W
= \rho(Z - N - 4 Z \bar x)$, $r \equiv \Delta Q_W / \bar Q_W$ is a function of
$\bar x$ alone; the $\rho$ dependence cancels.  The errors in $\bar x$ and $r$
are related to one another by
\beq
\frac{\delta r}{r} \approx \frac{4Z}{Z - \bar N - 4Z \bar x} \delta \bar x~~~,
\eeq
where $\bar N \equiv (N_1+N_2)/2$. For $^{133}_{55}{\rm Cs}$, the coefficient
is $4Z/(Z - \bar N -4Z \bar x) \approx - 3$, so that in order to obtain a
measurement of $\bar x$ to $\pm 0.0004$ (competitive with the average of the
LEP and SLC determinations mentioned in Table 1), one must measure $r$ to 0.1\%
of its value.  (This is considerably more demanding than requiring the
isotope {\it ratio} $Q_W(N_1)/Q_W(N_2) = 1 + \Delta Q_W/Q_W(N_2)$ to be
measured to 0.1\%.)  At this level it is likely that isotope-dependent
effects and uncertainties in electroweak radiative corrections become
significant.  Some statistical power can be added to the determination of
$\Delta Q_W$ if more than two isotopes are used.

\begin{figure}
\centerline{\epsfysize = 4in \epsffile{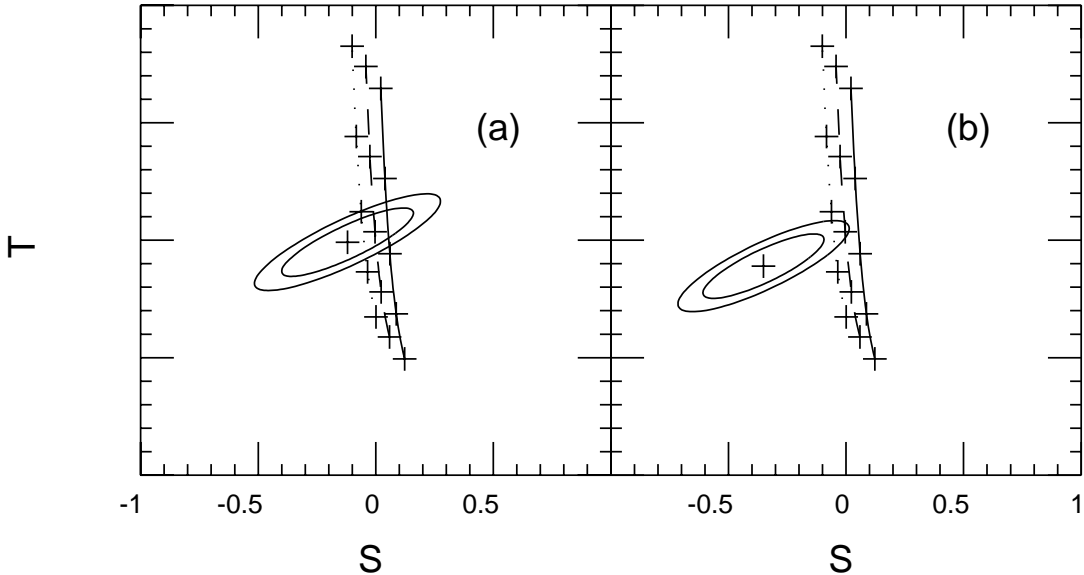}}
\caption{Allowed ranges of $S$ and $T$ at 68\% (inner ellipses) and 90\%
(outer ellipses) confidence levels.  Dotted, dashed, and solid lines correspond
to standard model predictions for $M_H = 100$, 300, 1000 GeV.  Tick marks, from
bottom to top, denote predictions for $m_t = 100$, 140, 180, 220, and 260 GeV.
(a) Fit including APV experiments with present errors; (b) errors on APV
experiments reduced by a factor of 4, with present central values of $Q_W$
retained.}
\end{figure}

In fact, the range of variation of models for the neutron charge radius in lead
\cite{PFW} is equivalent to about a 1\% uncertainty in $\sin^2 \theta$,
comparable to that envisioned for several future experiments involving parity
violation in the scattering of medium-energy polarized electrons \cite{med} on
nucleons and nuclei.  A calculation of the uncertainty due to the neutron
charge radius in cesium \cite{CV} is more optimistic, corresponding to
an error of 0.5\% in $\sin^2 \theta$ for a measurement with $N_1 = 70,~N_2 =
84$.

The theoretical error in $r$ for cesium \cite{MR} is itself about 0.2\%, and is
dominated by the error in the coefficient of $\sin^2 \theta$ in $Q_W$; most of
the error cancels in $\Delta Q_W$.  Other determinations of $\sin^2 \theta$ at
low momentum transfers $|q^2| \ll M_Z^2$ include the most recent CHARM II
result \cite{CHARMII}, $\sin^2 \theta = 0.2324 \pm 0.0083$, a measurement to
3.6\% accuracy, and the ratio $R_\nu$ mentioned above, which is roughly
equivalent to a measurement of the on-shell parameter $\sin^2 \theta_W \equiv 1
- M_W^2/M_Z^2 = 0.225 \pm 0.005$.  An error $\delta \sin^2 \theta_W$ is
equivalent by virtue of (3) and (4) \cite{JRRC} to an error $\delta \hat s^2 =
(\hat s^2/\hat c^2) \delta \sin^2 \theta_W \approx 0.3 \delta \sin^2 \theta_W$,
where $\hat c^2 \equiv 1 - \hat s^2$.  Thus deep inelastic neutrino scattering
is now providing a measurement of $\sin^2 \theta$ at $|q^2| \ll M_Z^2$ to
slightly better than a percent, but with residual dependence on top quark and
Higgs boson masses.

We conclude with a comparison of absolute and relative measurements of $Q_W$
for discovering or placing limits on effects of new gauge bosons.  In
Ref.~\cite{MR} the effect of a $Z_\chi$ (the extra $Z$ in SO(10) theories) was
expressed as
\beq \label{eqn:dq}
\Delta Q_{W~\rm tree}^{\rm new} \simeq 0.4 (2N + Z)(M_W/M_{Z_\chi})^2~~~.
\eeq
The central value of $\Delta Q_W  = (-71.04 \pm 1.81) - (-73.20 \pm 0.13) =
2.16 \pm 1.81$ in cesium, with $N = 78$ and $Z = 55$, could be accounted for
with a $Z_\chi$ of mass 500 GeV, to be compared with the lower
bound of 425 GeV set by a direct search at the Tevatron \cite{CDFZ}.  Thus, to
place a bound $M_{Z_\chi} > 1$ TeV, one would have to reduce the discrepancy to
$\Delta Q_W < 0.54$, requiring about a factor of four greater accuracy than the
present determination.

To obtain a bound $M_{Z_\chi} > 1$ TeV by measuring an isotope ratio, one would
have to measure $r$ in cesium to 0.2\%.  To see this, we express
\beq
r = \frac{\Delta N [-\rho + 0.8 (M_W/M_{Z_\chi})^2]}
{\rho(Z - \bar N - 4 Z \sin^2 \theta) + 0.4(2 \bar N + Z)(M_W/M_{Z_\chi})^2}~~,
\eeq
where $\Delta N \equiv N_1 - N_2$.  Expanding to first order in ${\cal R}
\equiv (M_W/M_{Z_\chi})^2/\rho$, we find
\beq
r \approx r^0 \left[ 1 + \frac{0.4Z(8 \sin^2 \theta - 3)}{\bar Q_W^0} {\cal R}
\right]~~~,
\eeq
where quantities with the superscript zero refer to those in the absence of the
$Z_\chi$ contribution.  Note that the terms with $\bar N$ cancel.  The
coefficient of ${\cal R}$ is about 0.34 for cesium and 0.31 for lead.  Thus, to
set a limit $M_{Z_\chi} > 1$ TeV, corresponding to ${\cal R} < 0.64\%$ with
$\rho \simeq 1$, one has to measure $r$ to 0.2\%.

The $Z_\chi$ is one of a family of possibilities arising in ${\rm E}_6$
theories, which also contain a boson $Z_\psi$ which arises when ${\rm E}_6$
breaks down to SO(10).  Let us parametrize a general $Z_\phi \equiv Z_\psi \cos
\phi + Z_\chi \sin \phi$.  Here $\phi$ is the same as the angle $\theta$
employed in Ref.~\cite{LR}, and opposite to the angle $\theta_U$ defined in
Ref.~\cite{SP}.  The boson sometimes called $Z_\eta$, which arises in
superstring theories, corresponds to $\phi = \arctan (3/5)^{1/2} \simeq
37.8^{\circ}$ in our notation.  We find that Eq.~(\ref{eqn:dq}) is merely
multiplied by a factor $f(\phi) \equiv \sin \phi [\sin \phi - (5/3)^{1/2} \cos
\phi]$.  This function vanishes at $\phi = 0$ and $\phi = 52.2^{\circ}$ and is
negative in between, attaining its most negative value of $-0.32$ at $\phi =
26.1^{\circ}$ and its maximum value of 1.32 at $\phi = 116.1^{\circ}$. The
corresponding bounds on $Z_\phi$ masses can be rescaled accordingly.

To summarize, atomic parity violation experiments can still play a key role in
providing information on fundamental parameters in particle physics, despite
recent strides in precise electroweak measurements. Absolute determination of
$Q_W$ for one or more atoms to an accuracy of half a percent is now the most
important goal.  This will help to constrain the Peskin-Takeuchi parameter $S$
in a useful manner and can roughly double the present lower limits on extra
gauge bosons. Measurements of ratios of isotopes are likely to provide
information on $\sin^2 \theta$ at low momentum transfers to an accuracy of at
best a percent, given present theoretical uncertainties about nuclear effects
in lead \cite{PFW}, or slightly better in cesium on the basis of the estimate
of Ref.~\cite{CV}. An error of a percent in $\sin^2 \theta$ is comparable to
that envisioned for other medium- and low-energy tests; indeed, deep inelastic
neutrino scattering already is close to providing such a constraint.
Measurement of the parameter $r \equiv \Delta Q_W/\bar Q_W$ to an accuracy of
about 0.5\% would constrain $\sin^2 \theta$ to a percent, while an accuracy of
0.2\% in $r$ would roughly double the present limit on new gauge boson masses.

I am indebted to W. Marciano, S. Pollock, P. Vogel, L. Wilets, and L.
Wolfenstein for useful discussions. I wish to thank the Institute for Nuclear
Theory at the University of Washington for hospitality during this work, which
was supported in part by the United States Department of Energy under Grant No.
DE FG02 90ER40560.

\def \ajp#1#2#3{Am. J. Phys. {\bf#1}, #2 (#3)}
\def \apny#1#2#3{Ann. Phys. (N.Y.) {\bf#1}, #2 (#3)}
\def \app#1#2#3{Acta Phys. Polonica {\bf#1}, #2 (#3)}
\def \arnps#1#2#3{Ann. Rev. Nucl. Part. Sci. {\bf#1}, #2 (#3)}
\def \cmts#1#2#3{Comments on Nucl. Part. Phys. {\bf#1}, #2 (#3)}
\def \cn{Collaboration}
\def \cp89{{\it CP Violation,} edited by C. Jarlskog (World Scientific,
Singapore, 1989)}
\def \efi{Enrico Fermi Institute Report No. EFI}
\def \f79{{\it Proceedings of the 1979 International Symposium on Lepton and
Photon Interactions at High Energies,} Fermilab, August 23-29, 1979, ed. by
T. B. W. Kirk and H. D. I. Abarbanel (Fermi National Accelerator Laboratory,
Batavia, IL, 1979}
\def \hb87{{\it Proceeding of the 1987 International Symposium on Lepton and
Photon Interactions at High Energies,} Hamburg, 1987, ed. by W. Bartel
and R. R\"uckl (Nucl. Phys. B, Proc. Suppl., vol. 3) (North-Holland,
Amsterdam, 1988)}
\def \ib{{\it ibid.}~}
\def \ibj#1#2#3{~{\bf#1}, #2 (#3)}
\def \ichep72{{\it Proceedings of the XVI International Conference on High
Energy Physics}, Chicago and Batavia, Illinois, Sept. 6 -- 13, 1972,
edited by J. D. Jackson, A. Roberts, and R. Donaldson (Fermilab, Batavia,
IL, 1972)}
\def \ijmpa#1#2#3{Int. J. Mod. Phys. A {\bf#1}, #2 (#3)}
\def \ite{{\it et al.}}
\def \jpb#1#2#3{J.~Phys.~B~{\bf#1}, #2 (#3)}
\def \lkl87{{\it Selected Topics in Electroweak Interactions} (Proceedings of
the Second Lake Louise Institute on New Frontiers in Particle Physics, 15 --
21 February, 1987), edited by J. M. Cameron \ite~(World Scientific, Singapore,
1987)}
\def \ky85{{\it Proceedings of the International Symposium on Lepton and
Photon Interactions at High Energy,} Kyoto, Aug.~19-24, 1985, edited by M.
Konuma and K. Takahashi (Kyoto Univ., Kyoto, 1985)}
\def \mpla#1#2#3{Mod. Phys. Lett. A {\bf#1}, #2 (#3)}
\def \nc#1#2#3{Nuovo Cim. {\bf#1}, #2 (#3)}
\def \np#1#2#3{Nucl. Phys. {\bf#1}, #2 (#3)}
\def \pisma#1#2#3#4{Pis'ma Zh. Eksp. Teor. Fiz. {\bf#1}, #2 (#3) [JETP Lett.
{\bf#1}, #4 (#3)]}
\def \pl#1#2#3{Phys. Lett. {\bf#1}, #2 (#3)}
\def \pla#1#2#3{Phys. Lett. A {\bf#1}, #2 (#3)}
\def \plb#1#2#3{Phys. Lett. B {\bf#1}, #2 (#3)}
\def \pr#1#2#3{Phys. Rev. {\bf#1}, #2 (#3)}
\def \prc#1#2#3{Phys. Rev. C {\bf#1}, #2 (#3)}
\def \prd#1#2#3{Phys. Rev. D {\bf#1}, #2 (#3)}
\def \prl#1#2#3{Phys. Rev. Lett. {\bf#1}, #2 (#3)}
\def \prp#1#2#3{Phys. Rep. {\bf#1}, #2 (#3)}
\def \ptp#1#2#3{Prog. Theor. Phys. {\bf#1}, #2 (#3)}
\def \rmp#1#2#3{Rev. Mod. Phys. {\bf#1}, #2 (#3)}
\def \rp#1{~~~~~\ldots\ldots{\rm rp~}{#1}~~~~~}
\def \si90{25th International Conference on High Energy Physics, Singapore,
Aug. 2-8, 1990}
\def \slc87{{\it Proceedings of the Salt Lake City Meeting} (Division of
Particles and Fields, American Physical Society, Salt Lake City, Utah, 1987),
ed. by C. DeTar and J. S. Ball (World Scientific, Singapore, 1987)}
\def \slac89{{\it Proceedings of the XIVth International Symposium on
Lepton and Photon Interactions,} Stanford, California, 1989, edited by M.
Riordan (World Scientific, Singapore, 1990)}
\def \smass82{{\it Proceedings of the 1982 DPF Summer Study on Elementary
Particle Physics and Future Facilities}, Snowmass, Colorado, edited by R.
Donaldson, R. Gustafson, and F. Paige (World Scientific, Singapore, 1982)}
\def \smass90{{\it Research Directions for the Decade} (Proceedings of the
1990 Summer Study on High Energy Physics, June 25--July 13, Snowmass,
Colorado),
edited by E. L. Berger (World Scientific, Singapore, 1992)}
\def \tasi90{{\it Testing the Standard Model} (Proceedings of the 1990
Theoretical Advanced Study Institute in Elementary Particle Physics, Boulder,
Colorado, 3--27 June, 1990), edited by M. Cveti\v{c} and P. Langacker
(World Scientific, Singapore, 1991)}
\def \yaf#1#2#3#4{Yad. Fiz. {\bf#1}, #2 (#3) [Sov. J. Nucl. Phys. {\bf #1},
#4 (#3)]}
\def \zhetf#1#2#3#4#5#6{Zh. Eksp. Teor. Fiz. {\bf #1}, #2 (#3) [Sov. Phys. -
JETP {\bf #4}, #5 (#6)]}
\def \zpc#1#2#3{Zeit. Phys. C {\bf#1}, #2 (#3)}
\def \zpd#1#2#3{Zeit. Phys. D {\bf#1}, #2 (#3)}

\end{document}